# Commensurability Effects in Viscosity of Nanoconfined Water

Mehdi Neek-Amal[1], Francois M. Peeters[2], Irina V. Grigorieva[3] and Andre K. Geim[3]

[1]Department of physics, Shahid Rajaee Teacher Training University, Lavizan, 16785-136, Teharn, Iran.

[2]Departement Fysica, Universiteit Antwerpen, Groenenborgerlaan 171, B-2020 Antwerpen, Belgium.

[3]School of Physics and Astronomy, University of Manchester, Manchester M13 9PL, UK.

## Abstract

The rate of water flow through hydrophobic nanocapillaries is greatly enhanced as compared to that expected from macroscopic hydrodynamics. This phenomenon is usually described in terms of a relatively large slip length, which is in turn defined by such microscopic properties as the friction between water and capillary surfaces, and the viscosity of water. We show that the viscosity of water and, therefore, its flow rate are profoundly affected by the layered structure of confined water if the capillary size becomes less than 2 nm. To this end we study the structure and dynamics of water confined between two parallel graphene layers using equilibrium molecular dynamics simulations. We find that the shear viscosity is not only greatly enhanced for subnanometer capillaries, but also exhibits large oscillations that originate from commensurability between the capillary size and the size of water molecules. Such oscillating behavior of viscosity and, consequently, the slip length should be taken into account in designing and studying graphene-based and similar membranes for desalination and filtration.

Water confined in a nanoscale channel exhibits very different properties compared to bulk water and its behavior depends strongly on the channel size and the affinity between water and a material of the confining walls.[1-8] Experimentally, large enhancement of the water flow rate was found for carbon nanotubes (CNT) with diameters less than a few nm[1-3] and, more recently, superpermeation of water through nanochannels within graphene membranes was reported.[9] The measured water flow in subnanometer CNTs can be several orders of magnitude higher than the values calculated from continuum hydrodynamic models2, which offers a possibility of developing new nanofluidics devices.

Numerous studies based on molecular dynamics (MD) simulations identified key factors that affect the behavior of water on the nanoscale[10-15]: (i) the breakdown of the uniform fluid density (layered structure of nanoconfined water); (ii) the water-solid wall slip length much larger than capillary sizes, which effectively results in different boundary conditions compared to bulk water in macroscopic channels, and (iii) the effective shear viscosity that is different from the bulk value.

In terms of the structure of water, there is consensus that nanoconfined water can form distinct ice-like layers at room temperature, including monolayer- and bilayer ice; different amorphous/ordered ice structures were found in a large number of MD studies. For example, monolayer ice was proposed by Zangi and Mark[12] using MD simulations with a five site and tetrahedrally coordinated model (TIP5P). They confined water between two parallel plates, applied a high lateral pressure ($P_l$) up to 2 kbar and found a non-flat monolayer of ice. Formation of a monolayer ice was also proposed as the reason for the superpermeation of water through graphene nanochannels[9,16].

Unlike the static structure, dynamics of water on nanoscale remains highly controversial, with widely varying and conflicting results (see Ref. (1) for a recent review). This is likely because boundary conditions are difficult to control, and measuring of the water flow close to a surface is difficult even in numerical simulations. The reported values of the slip length ($L_s$) for hydrophobic surfaces - a crucial parameter for determining water flow - vary by several orders of magnitude.[1] For example, using nonequilibrium MD simulations Kotsalis *et al.*[7] found $L_s$ for water flow in a 2.71nm diameter CNT to be 11 nm, while Ming D. Ma *et al.*[17], reported the slip length of about 200-1100 nm using the same method and a nanotube of the same dimensions. It was also reported that $L_s$ is sensitive to the external field (a constant force applied to all atoms within water molecules in order to force the flow of water through CNTs).[1,10,11] The situation is rather similar for a flat graphene, in which case $L_s$ is found to be scattered in the range 1 to 80 nm.[1,11]

In its turn, the slip length is a function of both the friction coefficient ($\xi$) and the viscosity of confined water ($\eta$) so that $L_s = \eta/\xi$.[1,10] The low-friction behavior of water inside hydrophobic capillaries is well understood as it is entropically unfavorable for a hydrophobic surface to bind to water molecules via ionic or hydrogen bonds, i.e., $\xi \propto \rho h$, where $\rho$ is the water density and $h$ is the distance between the two surfaces.[1] Furthermore, the friction is affected by the surface curvature[18] and depends on the atomic structure of the confining walls, e.g. $\xi$ for water on an ionic surface (hBN) is found to be three times larger than on graphene.[19]

In contrast, the shear viscosity of nanoconfined water is understood poorly, especially for channels narrower than a few nm.[20] Most of the experiments measured the viscosity of a liquid (e.g. water) between mica surfaces.[21-25] Raviv et al.[24,25] found that the effective viscosity of water remains within a factor of three of its bulk value when confined to films with thickness between 0.4 and 3.5 nm, while Dhinojwala et al.[26] found the viscosity under confinement orders of magnitude larger than in the bulk. On the other hand, MD simulations found monotonically decreasing (vanishing) viscosity of water inside CNTs as the CNT diameter falls below 1.5 nm.[27,28]

One of possible reasons for the conflicting results is the fact that the existing studies do not take into account that the viscosity of liquids near solid surfaces is spatially varying which should affect the effective viscosity of water in a nanochannel and, as a result, will affect the slip length and the flow rate. Yet, in the simulations, the channel size is often varied in ~1 nm steps. Therefore the effect of sub-Angstrom changes in the channel size on the viscosity and slip length, while probably important, remains unknown.

Here, by using the reactive force field (ReaxFF) potential[29] and employing equilibrium molecular dynamics simulations, we investigate the effect of the channel size on the structure and shear viscosity of water confined between two graphene layers. We show that sub-Angstrom variations in the channel size change the shear viscosity of water in an unexpected way. In particular, we find that the shear viscosity oscillates as a function of the distance between the confining walls, which originates from the interplay between the molecular size of water, hydrophobicity of the walls, and the complex arrangement of hydrogen bonds. Such oscillations should have a profound effect on the slip length and the flow rate and should be taken into account in studies of water transport through nanochannels.

## The model

There are two widely used molecular dynamics methods in water transport simulations: nonequilibrium molecular dynamics (NEMD) and equilibrium molecular dynamics (EMD) simulations using classical force fields, *e.g.* TIP4P and SPC/E models.[12,13] The simulation setup in a typical NEMD simulation is similar to that of a real experiment[30]: a field (pressure or density gradient) is applied at both ends of the capillary, then by analyzing the velocity profile the slip length is found by using

$$L_s = \frac{v_x}{\frac{\partial v_x}{\partial z}}|_{z=0}. \qquad (1)$$

The results from the NEMD simulations are very sensitive to the applied field as well as to averaging processes and may cause some unexpected errors.[1] In a typical EMD simulation, the friction coefficient is determined using velocity autocorrelation function and velocity-force correlation function.[31] The EMD method nowadays is commonly used which overcomes the boundary conditions and field dependence problems in NEMD.[1]

We employed EMD simulations using the ReaxFF[29] potential as implemented in the wellknown large scale atomic/molecular massively parallel simulator LAMMPS.[32] Note that the ReaxFF potential accounts for different possible bond formation and bond dissociation of different bond orders. It also contains Coulomb and van der Waals potentials to describe non-bonding interactions between all atoms. One of the main advantages of ReaxFF is that it calculates the charge polarization within the molecules which is achieved by using electronegativity and hardness parameters based on the electronegativity equalization and charge equilibration methods. Furthermore, the ReaxFF potential allows bond extension/contraction in water as well as bond-angle bending and it allows charge relaxation over each atom. This is in contrast to traditional force fields for water, *e.g.* SPC and TIP4P[33] (a rigid planar four site interaction potential for water) that keep the water molecules rigid during the MD simulations.

In the linear transport regime EMD simulations also can be used to give the viscosity for a liquid. It can be obtained from pressure fluctuations using the Green-Kubo equation[38]:

$$\eta_s = \lim_{t \to \infty} \frac{V}{k_B T} \int_0^t \sum_{\alpha,\beta} \frac{1}{5} \langle\langle P_{\alpha,\beta}(t) P_{\alpha,\beta}(0) \rangle\rangle, \tag{2}$$

where the averaging refers to thermal averaging and averaging over all water molecules, V is the volume, $T$ the temperature of the system, and $k_B$ the Boltzmann constant. The integrand is the autocorrelation function of the pressure tensor $P_{i,j}$ that has the following five independent components:
$P_{xx}$- $P_{yy}$ , $P_{yy}$- $P_{zz}$, $P_{xy}$, $P_{xz}$, and $P_{yz}$. The two summations are taken over all five terms. The autocorrelation function for measuring the shear viscosity of water decays fast over a short period of time *i.e.* of the order of a few ps.[39] We saved the $P_{i,j}$ values every 1 fs and averaged over a time interval of 5 ps (for more details see section "Method").

Results
Shear viscosity of bulk water

In order to check the validity of the ReaxFF potential for the simulation of the viscosity of water, we first performed extensive simulations for N=5488 water molecules in a box with dimensions 88×88×42.5 Å³ which corresponds to the density of water at room temperature, ≃1g/cm³. We applied periodic boundary conditions in the three directions and kept the temperature fixed at 298K using Nose-Hoover thermostat. After reaching the equilibrium state, we calculate the shear viscosity using Eq. (2). The viscosity is found to be (4.731±0.003)×10⁻⁴ Pa s which is in agreement with the results obtained from other force fields, *e.g.* TIP4P yields (4.83±0.09)×10⁻⁴ Pa s.[39] The experimental viscosity for bulk water is ≃6.0×10⁻⁴ Pa s[40] at room temperature. The difference between experiment and MD results is likely due to the relatively small unit cell in our MD simulations.

## Shear viscosity of nanoconfined water

Next we confine water between two graphene layers while keeping the density of water molecules fixed at $1 g/cm^3$. The density is given by $\rho = \frac{18N}{AN_a(h-t)}$ $g/cm^3$ for a fixed area $A=43\times37.5$ Å$^2$ (here $N_a$ is Avogadro's constant). Accordingly, we change the distance between the graphene layers from $h=7.5$Å to $h=20$Å. We note that, when estimating the volume of water, one needs to exclude the volume taken up by the graphene layers by using $h-t$ instead of $h$. In our calculations we used $t\sim1$Å as an approximation for the atomic diameter of carbon atoms.

In Fig. 1, we show the variation of the shear viscosity with $h$ at room temperature. It is seen that the shear viscosity increases for $h$ below ~18Å and is much higher than for bulk water either at ambient or elevated pressures[40]- see the horizontal dashed lines in Fig. 1. The highest viscosity is found for $h=7.5$Å. As $h$ increases, the viscosity first decays fast and then starts oscillating as a function of $h$. We have found four minima located at $h=9$, 11.5, 15 and 18Å and four maxima located at $h=7.5$, 10, 13.5, and 16.5Å.

In order to interpret the obtained results, we divide Fig. 1 into five segments shown by different colors and will analyze the structure of water at each minimum and maximum.

## Structures

In agreement with previous studies[12,16] we found that water confined in the graphene capillary forms distinct layers. In Fig. 2, we show four snapshots of the cross-sectional view of the structure of confined water corresponding to the maxima (a) and minima (b) of the shear viscosity. To further characterize the degree of ordering into layered structures, Figs. 3(a) and (b) show the corresponding radial distribution function (RDF) for the O-O distances (for all water molecules) for the channel sizes corresponding to maxima and minima in shear viscosity, respectively. Furthermore the corresponding density profiles along the perpendicular direction are shown in Fig. 4.

*Maxima of the shear viscosity.*

For the narrowest channel (*i.e.* $h$=7.5Å, the first maximum in the viscosity) a typical snapshot of the water structure is shown in Fig. 2(a) (top panel). Notice that two distinct layers of water are formed between the graphene walls, although no ordering was found in the arrangement of water molecules within the layers. Furthermore, the layered structure is very stable: there is no exchange of water molecules between the two layers. The corresponding RDF (circles in Fig. 3(a)) has two clear peaks at 2.84 and 5.4Å. The first peak corresponds to the nearest neighbor average O-O distance and the second peak indicates a long range order in the system.

For the 10Å channel, the first and second peaks in Fig. 3(a) are at 2.84Å and 5.7Å, respectively. Note that the second peaks for $h$=7.5Å and 10Å are closer to each other both in height and position than to the (less pronounced) peaks for $h$=15Å and 16.5Å. Furthermore, the RDF values for these $h$ are notably larger than the corresponding RDF for $h$=13.5Å and 16.5Å. This indicates a similar long range order for $h$=7.5Å and 10Å, with next-nearest neighbor O-O distance of 5.5Å while the distinctiveness of the water layers and the long range order are gradually lost for wider channels.

In Fig. 4(a), we show the density profile of the water molecules for the channels with $h$=7.5Å and 10Å. There are three water layers in the $h$=10Å channel, each layer corresponding to a peak in the density profile in Fig. 4(a). Importantly, while for $h$=7.5Å there is no exchange of water molecules between the layers (the structure is solid-like, *i.e.* corresponds to ice), for $h$=10Å, a 'virtual' layer forms with lower density than the 'main layers' adjacent to the graphene walls. The water molecules in this virtual layer are in constant exchange with the main layers. This is the reason that in Fig. 2(a) the middle layer for $h$=10Å looks more diffuse.

The shear viscosity in the $h$=7.5Å channel is almost 8 times higher than that for $h$=10Å. For $h$=13.5Å (16.5Å), the viscosity is further reduced by an order of magnitude. As seen form Fig. 4(b), for these wider channels there are four (five) distinguishable water layers with two denser, well-defined layers adjacent to the graphene walls and more diffuse intermediate layers with almost uniform distribution of water molecules, especially for $h$ =16.5Å.

*Minima of the shear viscosity.*

Water molecules in the channel with $h$=9Å form two layers adjacent to the graphene walls (main layers), somewhat similar to the $h$=7.5Å channel where the highest viscosity is observed. However, in this case, there are also many water molecules between the two layers that effectively connect the main layers with each other (see Figs. 2(b) and 4(c)). For the $h$=11.5Å channel such water molecules clearly form a middle layer. However, this middle layer is not stable: its water molecules constantly hop between the top and bottom layers (main layers). This movement of water between the layers can be seen in provided Supplementary Movies (see Supplementary information). We note that the number of water layers formed in the graphene channel for $h$=9Å and $h$=11.5Å is in agreement with the MD results in Ref.(16).

For $h$=15 and 18Å channels the middle layers become almost indistinct (see density profiles in Fig. 4(d)), again with water molecules in intermediate layers constantly hopping and connecting the two main layers adjacent to the graphene walls. It is this hopping of water molecules between the main layers that plays an essential role in decreasing the shear viscosity.

In fact, at the minima the viscosity of confined water becomes of the same order as for bulk water at ambient pressure[40] (the latter is shown in Fig. 1). This is consistent with the absence of a stable layered structure due to the constant exchange of water molecules between the layers (20% of water molecules hop between the main layers).

In order to compare the dynamic properties of water corresponding to, for example, the first maximum and the first minimum of the shear viscosity, we performed two long time simulations for $h$=7.5Å and $h$=9Å. The results for a time series of the variation of z-component of the center of mass of confined water molecules

$$Z_{cm}(t) = \frac{\sum_i^{N(h)} Z_i(t)}{N(h)} \qquad (3)$$

are shown in Fig. 5. The hopping between two main layers in the case $h$=9Å is seen as larger fluctuations in $Z_{cm}$ around $Z$=0. For $h$=7.5Å we found much smaller fluctuations around $Z$=0 which indicates freezing of water (see two corresponding movies in the Supplementary Information). The larger fluctuations of $Z_{cm}$ yield diffusive motion of water molecules along $Z$ which reduces the shear viscosity.

Discussion

Our molecular dynamics simulations reveal the strong sensitivity of the shear viscosity of confined water to the size of the confining channel. Increasing the channel size, even by as little as 1Å, results in changes of the shear viscosity by more than an order of magnitude. Distinct layering of water confined between the graphene layers is apparent only for particular channel sizes. We also find that i) independent of $h$, two ' main layers' of water are always formed, adjacent to the graphene walls (which also have the largest number of water molecules as compared to the intermediate layers), ii) as $h$ increases beyond 15-16Å, the middle layers become mixed, and we obtain almost bulk water between the two main water layers, and iii) the first peak in the O-O distance, corresponding to the nearest layer separation occurs around 2.84Å, almost independent of $h$, while the location of the second peak (next-nearest layer separation) depends on $h$.

From our analysis, it is clear that the shear viscosity of confined water is related to the commensurability between the size of the channel and the space required to accommodate one layer of water molecules. For $h$ values corresponding to low shear viscosity (of the same order as the shear viscosity of bulk water), e.g. for $h$=9 and 11.5Å, confined water is commensurate with the channel size $h$. The latter results in minima in the shear viscosity. For $h$ values corresponding to the larger shear viscosity, confined water is incommensurate with $h$, resulting in maxima of the shear viscosity. Here commensurability is controlled by two main parameters: density of water and the size of water molecules.

One would expect that the following equation is obeyed for the distance between the two graphene layers forming the channel filled with layered water:
$$h_i(n) = 2\delta + (n-1)d, \qquad (4)$$

where $d$ is the distance between the water layers, $n$ is the number of water layers, and $\delta$ is the distance between the graphene sheets and one of the adjacent water layers. We would expect that in general (ignoring the incommensurate effects and for a constant water density) $h_i$ should be only a function of '$n$' and consequently $\delta \cong \delta_0 \cong 2.7$Å should be independent of $h$. In order to find $\delta$ and $d$, we performed several annealing MD simulations by cooling the system down to zero Kelvin and were able to calculate $\delta$

| $h$(Å) | $d$(Å) | $\delta$(Å) |
|---|---|---|
| 7.5 | 2.4±0.1 | 2.5±0.1 |
| 10 | 2.4±0.1 | 2.6±0.1 |
| 9 | 3.6±0.2 | 2.7±0.2 |
| 11.5 | 3.0±0.2 | 2.7±0.3 |

TABLE I. The distance between water layers, *i.e. d*, and the distance between the two main water layers and the graphene walls, *i.e. $\delta$*.

and $d$ which are listed in Table I. In contrast to expectations, the obtained $d$ (and calculated $\delta = \frac{h-(n-1)d}{2}$), for the two first maxima and minima indicate that $\delta$ and $d$ are a function of $h$. We found that $\delta \approx \delta_0$ is only valid for the minima of the shear viscosity, *i.e.* $h$=9Å and 11.5Å, giving the condition

*Commensurate case:* $\quad \frac{d\eta_s}{dh} = 0, \quad \frac{d\eta_s}{dh} > 0.$ (5)

For $h$=7.5 and 10Å, we found $\delta < \delta_0$ which refers to the phase of water with closer layers (shorter $d$ with respect to the minima cases) and all water molecules strongly localized within the layers. The corresponding condition is

*Incommensurate case:* $\quad \frac{d\eta_s}{dh} = 0, \quad \frac{d\eta_s}{dh} < 0.$ (6)

With decreasing $d$ and $\delta$ the shear viscosity increases leading to oscillations between commensurate cases (*e.g.* h=9Å and 11.5Å) and incommensurate cases (*e.g.* h=7.5Å and 10Å). The smaller the value of $d$, the larger the vdW force between the water layers and consequently the larger the shear viscosity.

It is interesting to note that the atomic force microscopy (AFM) studies show oscillations in solvation force as the distance between the AFM tip and the surface was varied, with a period approximately equal to the molecular size.[41] The layering of confined liquid and the formation and breaking of hydrogen bonds between the tip and the surface are the main reasons for these oscillations.[20] Cleveland *et al.*[42] observed hopping of the hydrophilic tip between four discrete levels with an average spacing of 2Å. They also observed several minima in the potential with spacing varying between 1.5 and 3.0Å, which are comparable to the size of a water molecule. It can be speculated that the force minima and force maxima are due to the formation of liquid-like and a solid-like structure of water, respectively, as found in our work. A solid behavior of the confined water between tip and a gold surface was reported in scanning tunneling microscopy measurements when the tip-surface distance is $h$=7Å[43], which is consistent with our finding for $h$=7.5Å case.

Conclusions

In summary, we show that the shear viscosity of confined water is controlled by the interplay between the spacing of the water layers and the width of the nanochannel: The viscosity oscillates with $h$ in such a way that it is minimum for the commensurate cases (where $\delta \geq \delta_0$ and and $d$ >3Å) and it reaches a maximum for the incommensurate cases (where $\delta < \delta_0$ and $d$ <3Å). This is in contrast to earlier MD simulations (*e.g.*, Refs.(27,28)) where monotonically decreasing viscosity was found for a decreasing channel size, presumably due to the different model used for water (TIP3P and TIP4P-EW, where the dipole/charge on each molrcule/atom is constant) and a different method for viscosity calculation (Eyring-MD method). As the channel size becomes larger than 2 nm, the shear viscosity decreases and approaches the viscosity of bulk water. This is in agreement with experiments that indicate that the viscosity of a liquid confined between two surfaces with $h$ >2-3nm remains close to that in the bulk.[21]

Our findings have an important implication that the slippage of water should also be strongly affected by the details of its molecular structure and the subnanometer changes in the channel size. Indeed, AFM measurements on a few-molecule thick layer of water confined between mica and an AFM tip found that dynamics of such water is determined predominantly by solvation effects that depend on the exact separation between the tip and the surface.[44]

Finally, it is worth mentioning that we performed several additional simulations for different densities of confined water (which varied between 1.7g/cm$^3$ and 0.7g/cm$^3$) and found the same decreasing and oscillating behavior of the shear viscosity. In these simulations we fixed the number of water molecules to N=2016 and changed the distance between the graphene layers. The shear viscosity for the highest density (1.7 g/cm$^3$ for $h$=7.5Å) is found to be close to 1.0 Pa s which is just twice that for water of density 1 g/cm$^3$ in a $h$=7.5Å channel shown in Fig. 2(a). Therefore the density of water can control its shear viscosity in a similar way as the channel size.

Method

The parameters used in our simulations to describe the interactions between carbon atoms in graphene and O and H atoms in water were taken from ffield.reax.FC source file in LAMMPS. Their suitability to model "-O-H, -C-O=H, C≡O", and *etc.* bonds has been validated in previous studies (*e.g.* Ref.(14,34-36). To further verify that these parameters are suitable for our study, we calculated the viscosity of bulk water (see below) and also used them to estimate the hydrogen bonding energy between water molecules. The latter yielded -0.16 eV/water molecule (-15.43 kJ mol$^{-1}$) which is in the same range as the known H-bond energy for bulk ice.[37]

The computational unit cell contains 1200 carbon atoms making up two rigid graphene layers that are separated by a distance $h$. In order to calculate the shear viscosity of water for a constant density, *i.e.* 1*g/cm*$^3$, we added N water molecules randomly (3N atoms) in the unit cell, where N should be a function of $h$, *i.e.* N($h$). Before performing the viscosity calculations we relax the system for 1.5 ps employing the NVT ensemble at 298K. In all simulations we apply periodic boundary conditions without applying any external lateral pressure and fix the temperature at 298 K. This allows us to study the effect of capillary size on the shear viscosity.

Supporting Information Available: Two movies showing molecular dynamics (MD) simulation results for the first minimum and the first maximum of the shear viscosity. This material is available *via* the Internet at http://pubs.acs.org.

Acknowledgment. M.N.A. was support by Shahid Rajaee Teacher Training University under contract number 29605.

## References

1. S. K. Kannam; B. D. Todd; J. S. Hansen; P. J. Daivis. How Fast Does Water Flow in Carbon Nanotubes? J. Chem. Phys. 2003, **138**, 094701-094709.

2. Jason K. Holt; Hyung Gyu Park; Yinmin Wang; Michael Stadermann; Alexander B. Artyukhin; Costas P. Grigoropoulos, Aleksandr Noy, and Olgica Bakajin. Fast Mass Transport Through Sub-2-Nanometer Carbon Nanotubes. Science 2006, **312**, 1034-1037.

3. M. Majumder; N. Chopra; R. Andrews; B. J. Hinds. Nanoscale Hydrodynamics: Enhanced Flow in Carbon Nanotubes. Nature (London) 2005, **438**, 44.

4. A. Kalra; S. Garde; G. Hummer. Osmotic Water Transport Through Carbon Nanotube Membranes. Proc. Natl. Acad. Sci. U.S.A. 2003, **100**, 10175.

5. R. Pit; H. Hervet; L. Leger. Direct Experimental Evidence of Slip in Hexadecane: Solid Interfaces. Phys. Rev. Lett. 2000, 85, 980-983.

6. G. M. Whitesides; A. D. Stroock. Flexible Methods for Microfluidics. Phys. Today 2001, **54**, 42-48.

7. E. M. Kotsalis; J. H. Walther; P. Koumoutsakos. Multiphase Water Flow Inside Carbon Nanotubes. Int. J. Multiphase Flow 2004, **30**, 995-1010.

8. G. Hummer; J. C. Rasaiah; J. P. Noworyta. Water Conduction Through the Hydrophobic Channel of a Carbon Nanotube. Nature (London) 2001, **414**, 188-190.

9. R. R. Nair; H. A. Wu; P. N. Jayaram; I. V. Grigorieva; A. K. Geim. Unimpeded Permeation of Water Through Helium-Leak-Tight Graphene-Based Membranes. Science 2012, **335**, 442-444.

10. Sridhar Kumar Kannam; B. D. Todd; J. S. Hansen; Peter J. Daivis. Slip Length of Water on Ggraphene: Limitations of Non-Equilibrium Molecular Dynamics Simulations. J. Chem. Phys. 2012, **136**, 024705-024709.

11. John. A. Thomas; Alan J. H. McGaughey. Reassessing Fast Water Transport Through Carbon Nanotubes. Nano Letters 2008, 8, 2788-2793.

12. R. Zangi; A. E. Mark. Monolayer Ice. Phys. Rev. Lett. 2003, **91**, 025502-025505.

13. R. Zangi; A. E. Mark. Electrofreezing of Confined Water. J. Chem. Phys 2004, **120**, 7123-7130.

14. M. Sobrino Fernandez Mario; M. Neek-Amal; F. M. Peeters. AA-stacked Bilayer Square Ice Between Graphene Layers. Phys. Rev. B 2015, **92**, 245428-245432.

15. Hu Qiu; Xiao Cheng Zeng; Wanlin Guo. Water in Inhomogeneous Nanoconfinement: Coexistence of Multilayered Liquid and Transition to Ice Nanoribbons. ACS Nano 2015, **9**, 9877-9884.

16. G. Algara-Siller; O. Lehtinen; F. C. Wang; R. R. Nair; U. Kaiser; H. A. Wu ; A. K. Geim; I. V. Grigorieva. Square Ice in Graphene Nanocapillaries. Nature (London), 2015, **519**, 443-445.

17. M. D. Ma; L. Shen; J. Sheridan; J. Z. Liu; C. Chen; and Q. Zheng. Friction of water slipping in carbon nanotubes. Phys. Rev. E, 2011, **83**, 036316-0.36322.

18. K. Falk; F. Sedlmeier; L. Joly; R. R. Netz; L. Bocquet. Molecular Origin of Fast Water Transport in Carbon Nanotube Membranes: Superlubricity *versus* Curvature Dependent Friction. Nano Lett. 2010, **10**, 4067-4073.

19. Gabriele Tocci; Laurent Joly; Angelos Michaelides. Friction of Water on Graphene and Hexagonal Boron Nitride from Ab Initio Methods: Very Different Slippage Despite Very Similar Interface Structures. Nano Letetrs 2014, **14**, 6872-6877.

20. A. Verdaguer; G. M. Sacha; H. Bluhm; M. Salmeron. Molecular Structure of Water at Interfaces: Wetting at the Nanometer Scale. Chem. Rev. 2006, **106**, 1478-1510.

21. J. N. Israelachvili. Measurment of the Viscosity of Liquids in Very Thin Films. J. Colloid Interface Sci. 1986, **110**, 263-271.

22. J. Klein; E. Kumacheva. Simple liquids confined to molecularly thin layers. I. Confinement-Induced Liquid-to-Solid Phase Transitions. J. Chem. Phys., 1998, **108**, 6996-7009.

23. J. N. Israelachvili; Gayle E. Adams. Measurement of Forces Between Two Mica Surfaces in Aqueous Electrolyte Solutions in the Range 0-100 nm . J. Chem. Soc., Faraday Trans. I, 1978, **74**, 975-1001.

24. U. Raviv; P. Laurat; J. Klein. Fluidity of Water Confined to Subnanometre Films. Nature (London), 2004, **413**, 51-54.

25. U. Raviv; J. Klein. Fluidity of Bound Hydration Layers. Science 2002, 297, 1540-1543.

26. A. Dhinojwala; S. Granick. Relaxation Time of Confined Aqueous Films Under Shear. J. Am. Chem. Soc. 1997, **119**, 241; Y. Zhu and S. Granick, Viscosity of Interfacial Water. Phys. Rev. Lett. 2001 **87**, 096104-096108.

27. Hongfei Ye; Hongwu Zhang; Zhongqiang Zhang; Yonggang Zheng. Size and Temperature Effects on the Viscosity of Water Inside Carbon Nanotubes. Nanoscale Research Letters, 2011, **6**, 87-91.

28. J. S. Babu; S. P. Sathian. The Role of Activation Energy and Reduced Viscosity on the Enhancement of Water Flow Through Carbon Nanotubes. J. Chem. Phys. 2011, **134**, 194509-194515.

29. A. C. T. van Duin; S. Dasgupta; F. Lorant; W. A. Goddard. ReaxFF: a Reactive Force Field for Hydrocarbons. J. Phys. Chem. A, 2001, **105**, 9396-9409.

30. David Schaeffel; Stoyan Yordanov; Marcus Schmelzeisen; Tetsuya Yamamoto; Michael Kappl; Roman Schmitz; Burkhard Dunweg; Hans-Jurgen Butt; and Kaloian Koynov. Hydrodynamic Boundary Condition of Water on Hydrophobic Surfaces. Phys. Rev. E, 2013, **87**, 051001-051004.

31. J. S. Hansen; B. D. Todd; P. J. Daivis. Prediction of Fluid Velocity Slip at Solid Surfaces. Phys. Rev. E, 2011, 84, 016313-016320.

32. S. Plimpton. Fast Parallel Algorithms for Short-Range Molecular Dynamics. J. Comp. Phys. 1995, **19**, 117-136.

33. J. L. Abascal; C. Vega. A General Purpose Model for the Condensed Phases of Water: TIP4P/2005. J. Chem. Phys. 2005 123, 234505-234516.

34. K Chenoweth; ACT Van Duin; WA Goddard. ReaxFF Reactive Force Field for Molecular Dynamics Simulations of Hydrocarbon Oxidation. J. Phys. Chem. A 2008, **112**, 1040-1053.

35. Jennifer L Achtyl, Raymond R Unocic; Lijun Xu; Yu Cai; Muralikrishna Raju; Weiwei Zhang; Robert L Sacci; Ivan V Vlassiouk; Pasquale F Fulvio; Panchapakesan Ganesh *et al.* Aqueous proton transfer across single-layer graphene. Nature Commun. **6** (2015)

36. J Yeon; SH Kim; ACT van Duin. Effects of Water on Tribochemical Wear of Silicon Oxide Interface: Molecular Dynamics (MD) Study with Reactive Force Field (ReaxFF). Langmuir 2016, **32**, 10181026.

37. M. Chaplin, arXiv:0706.1355.

38. Jean-Pierre Hansen; I. R. McDonald. *Theory of Simple Liquids With Applications to Soft Matter*, (ScienceDirect 2013).

39. George S. Fanourgakis; J. S. Medina; R. Prosmiti. Determining the Bulk Viscosity of Rigid Water Models. J. Phys. Chem. A, 2012, 116, 2564-2570.

40. Petra Först; Franz Werner; and Antonio Delgado. The Viscosity of Water at High Pressures-Especially at Subzero Degrees Centigrade. Rheol Acta 2000, **39**, 566-573.

41. Jianping Gao; W. D. Luedtke; Uzi Landman. Origins of Solvation Forces in Confined Films. J. Phys. Chem. B, 1997, 101, 4013-4023.

42. J. P. Cleveland; T. E. Schaffer; P. K. Hamsa. Probing Oscillatory Hydration Potentials Using Thermal-Mechanical Noise in an Aatomic-Force Microscope, Phys. Rev. B, 1995, **52**, R8692.

43. E.-M. Choi; Y.-H. Yoon; S. Lee; H. Kang. Freezing Transition of Interfacial Water at Room Temperature under Electric Fields. Phys. Rev. Lett. 2005, **95**, 085701-085703.

44. S. Jeffery, P. M. Hoffmann, J. B. Pethica, C. Ramanujan, H. O. Ozer, and A. Oral. Direct Measurement of Molecular Stiffness and Damping in Confined Water Layers. Phys. Rev. B, 2014, **70**, 054114-054121.

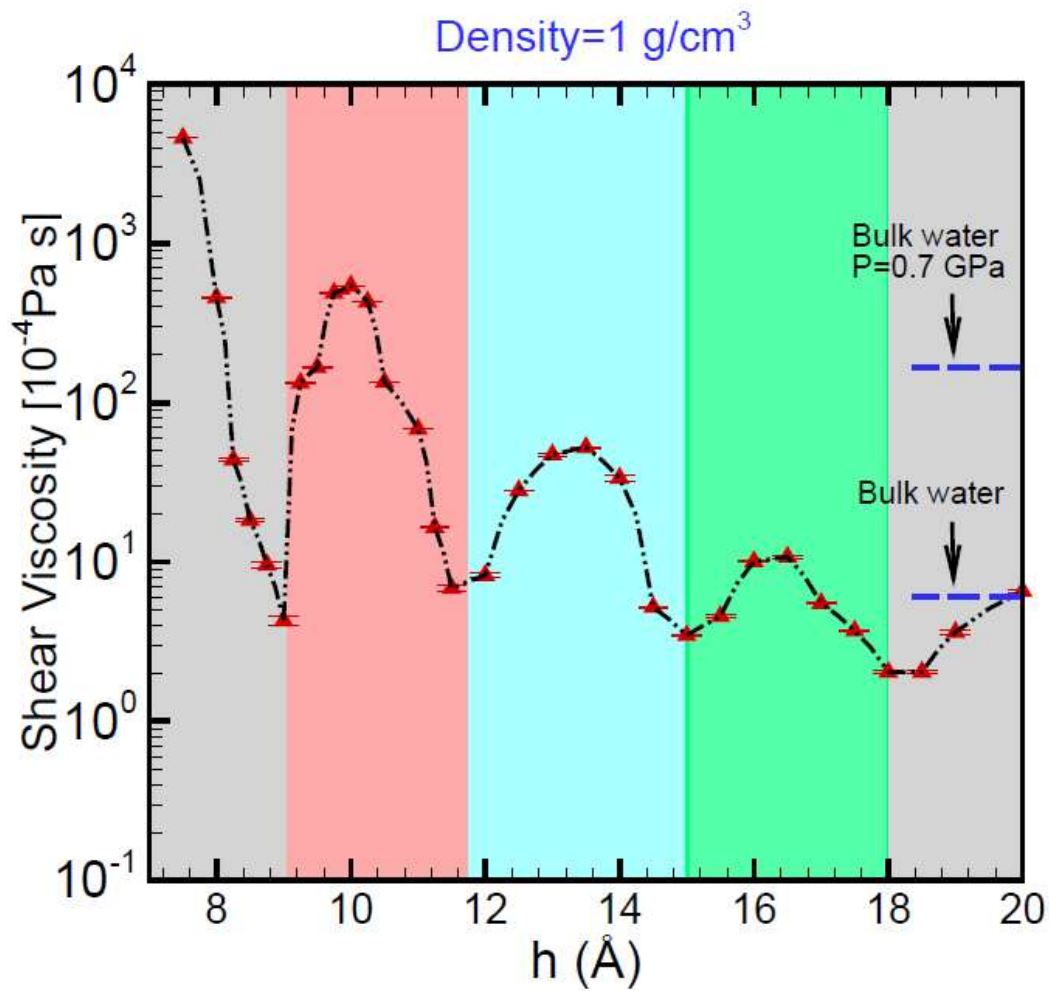

Figure 1. (color online) The variation of the shear viscosity of confined water between two graphene layers which are separated by a distance *h*. The result for bulk water is taken from Ref (40). The dashed line is a linear interpolation of the data.

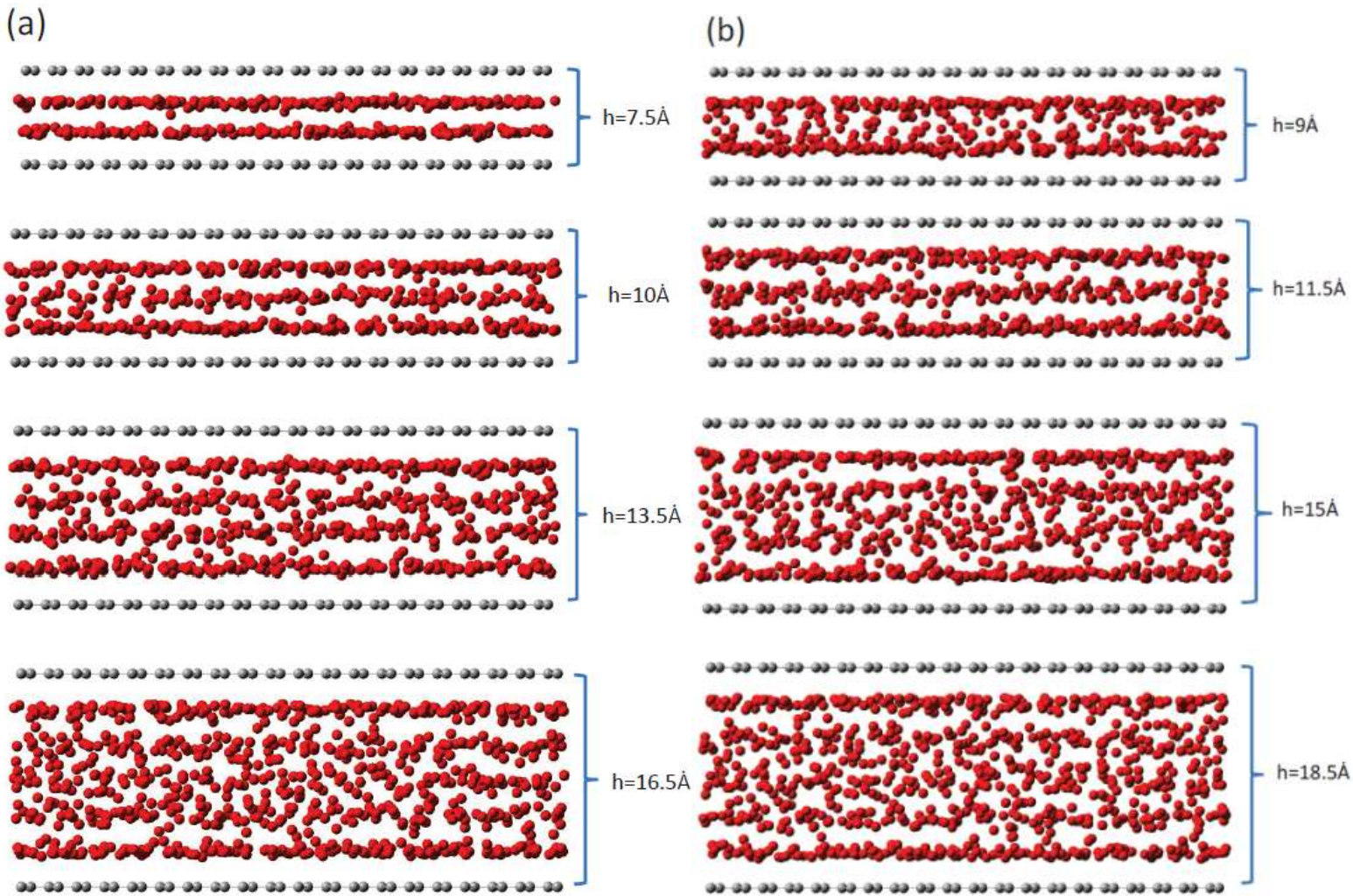

Figure 2. (color online) Side view of snapshots of water at T=298K confined between two graphene layers separated by different distances. The (a) and (b) figures show the results corresponding to the maxima and minima of the shear viscosity shown in Fig. 1, respectively.

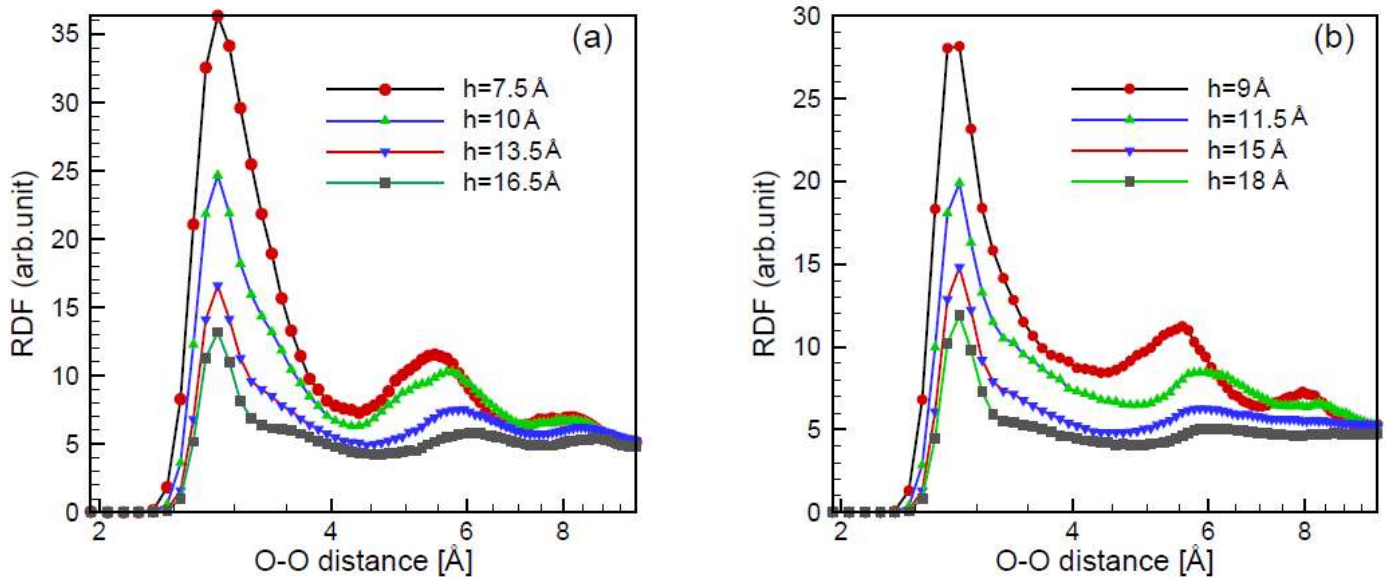

Figure 3. (color online) Radial distribution function of the positions of oxygen atoms for (a) maximum and (b) minimum values of the shear viscosity of confined water between the graphene layers. The O-O distances are shown in logarithmic scale for clarity.

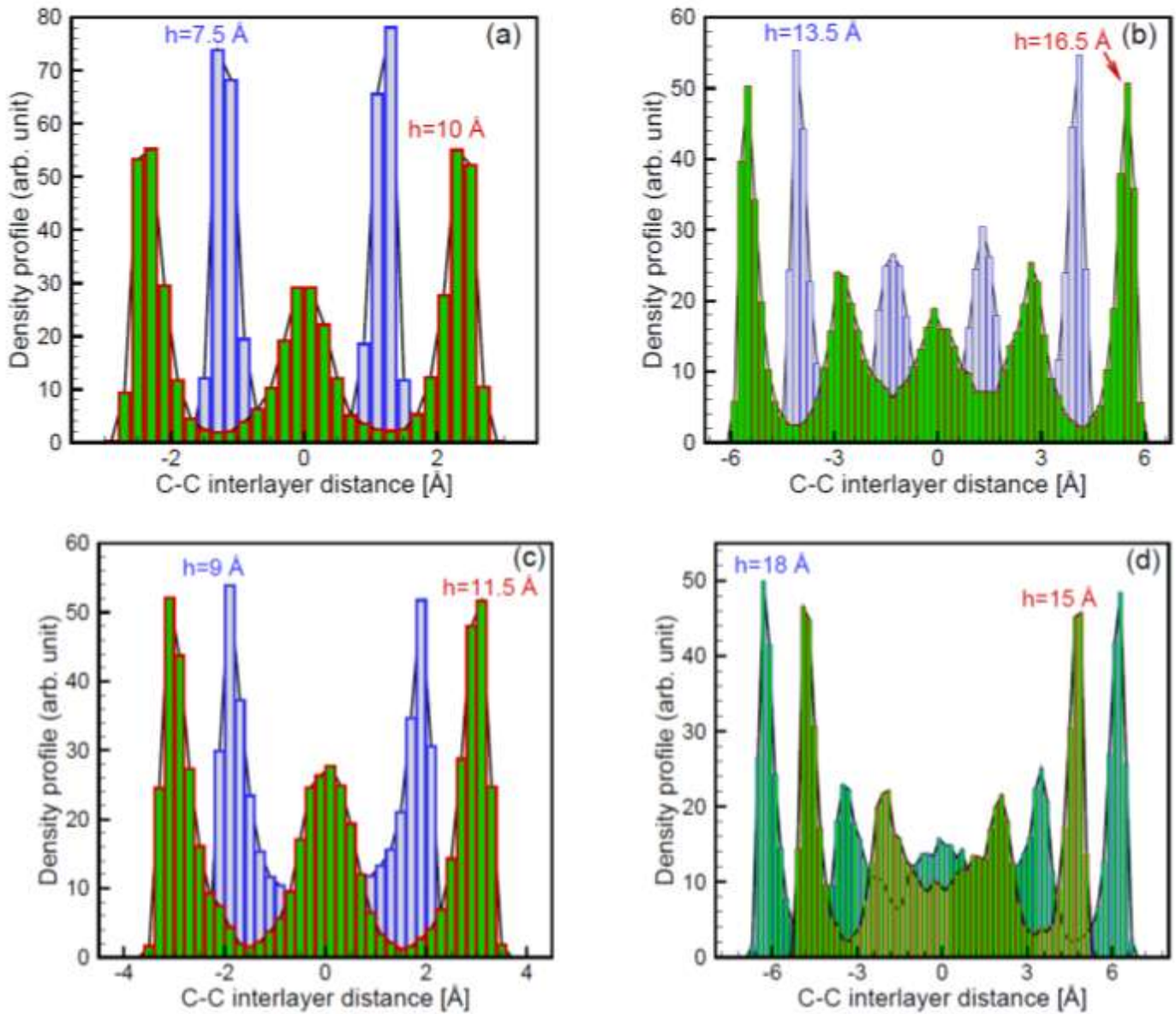

Figure 4. (color online) Density profiles (perpendicular to the graphene layers) for the snapshots shown in Fig. 2. (a,b) are related to the maxima and (c,d) are related to the minima of the shear viscosity.

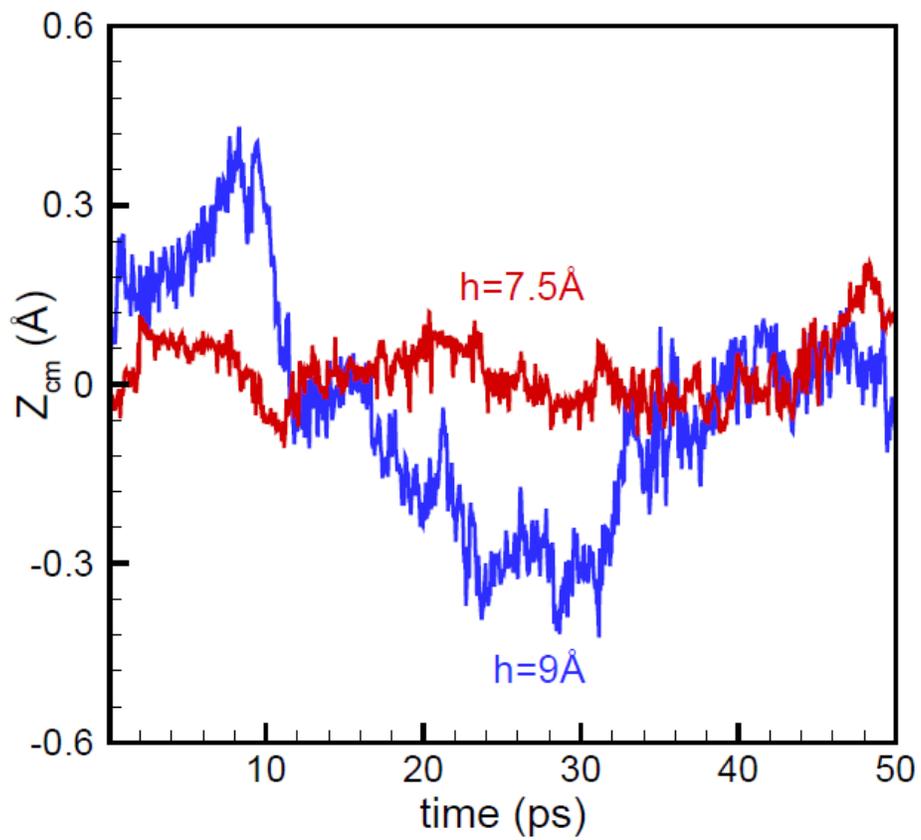

Figure 5. (color online) The time evolution of the z-component of the center of mass of confined water between two graphene layers which are separated by *h*=7.5Å and *h*=9Å.